\documentclass[prl,aps,twocolumn,nofootinbib,superscriptaddress,amsmath,amssymb]{revtex4-2}

\usepackage[T1]{fontenc}
\usepackage[utf8]{inputenc}
\usepackage{graphicx}
\usepackage{bm}
\usepackage{xcolor}
\usepackage[colorlinks=true,allcolors=green!40!black]{hyperref}

\newcounter{appcnt}
\renewcommand{\theappcnt}{\Alph{appcnt}}
\newcommand{\appsec}[2]{\refstepcounter{appcnt}%
  \subsection*{Appendix \theappcnt: #1}\label{#2}}

\newcounter{smcnt}
\renewcommand{\thesmcnt}{\Roman{smcnt}}
\newcommand{\smsec}[2]{\refstepcounter{smcnt}%
  \subsection*{\thesmcnt. #1}\label{#2}}
\newcommand{\smcite}[1]{\cite[Sec.~\ref{#1}]{SM}}

\newcounter{thmcnt}\newcounter{lemcnt}\newcounter{obscnt}\newcounter{defcnt}\newcounter{propcnt}
\newcommand{\Thm}[1]{\refstepcounter{thmcnt}\label{#1}Theorem~\thethmcnt}
\newcommand{\Lem}[1]{\refstepcounter{lemcnt}\label{#1}Lemma~\thelemcnt}
\newcommand{\Obs}[1]{\refstepcounter{obscnt}\label{#1}Observation~\theobscnt}
\newcommand{\Def}[1]{\refstepcounter{defcnt}\label{#1}Definition~\thedefcnt}
\newcommand{\Prop}[1]{\refstepcounter{propcnt}\label{#1}Proposition~\thepropcnt}

\newcommand{\beqa}{\begin{eqnarray}}
\newcommand{\eeqa}{\end{eqnarray}}
\newcommand{\beq}{\begin{equation}}
\newcommand{\eeq}{\end{equation}}
\newcommand{\Scr}{{\rm Scrooge}}                    % ensamble de Scrooge
\newcommand{\Haar}{{\rm Haar}}                      % medida de Haar
\newcommand{\tr}{{\rm tr}}                          % traza
\newcommand{\ket}[1]{\left \vert #1 \right \rangle}    % ket
\newcommand{\bra}[1]{\left \langle #1 \right \vert}    % bra
\newcommand{\ketbra}[2]{\ket{#1}\bra{#2}}             % ketbra

\begin{document}

\title{Maximum Rényi entropy principle for the Scrooge ensemble from local filtering}

\author{Agustín \surname{Agote}}
\affiliation{Universidad de Buenos Aires, Facultad de Ciencias Exactas y Naturales,
Departamento de Física, Ciudad Universitaria, 1428 Buenos Aires, Argentina}
\affiliation{CONICET - Universidad de Buenos Aires, Instituto de Física de Buenos Aires (IFIBA),
Ciudad Universitaria, 1428 Buenos Aires, Argentina}
\email[Contact author: ]{a.agote@df.uba.ar}
\author{Augusto J. \surname{Roncaglia}}
\affiliation{Universidad de Buenos Aires, Facultad de Ciencias Exactas y Naturales,
Departamento de Física, Ciudad Universitaria, 1428 Buenos Aires, Argentina}
\affiliation{CONICET - Universidad de Buenos Aires, Instituto de Física de Buenos Aires (IFIBA),
Ciudad Universitaria, 1428 Buenos Aires, Argentina}

\begin{abstract}
Measuring part of a chaotic many-body system leaves the rest in a pure state conditioned on the outcome. At fixed average state $\rho$, this projected ensemble approaches the Scrooge ensemble. We show that Scrooge is the unique maximizer, at fixed full-rank $\rho$, of the Rényi entropy of order $d/(d+ 1)$ relative to Haar, with $d$ the dimension of the unmeasured part. This entropy is the only separable one whose maximization commutes with postselected local filtering. When $\rho$ is the only constraint, this requirement fixes Scrooge. The same entropy is also the maximum Shannon entropy of the unnormalized post-measurement vectors, and the results extend to Scrooge analogues for mixed states, Gaussian fermionic states and systems with conserved charges.
\end{abstract}

\maketitle

\textit{Introduction}.--- Quantum simulators that record the full outcome of a measurement have led to a refined notion of thermalization, ``deep thermalization''~\cite{Cotler23,Choi23,Ho22,Ippoliti23,Mark26}. 
In this setting, one prepares a pure state $\ket\Psi$ of $A\otimes B$, with $A$ of dimension $d$, and measures $B$ in a basis $\{\ket b\}$. With probability $p_b$, system $A$ is left in the pure state $\ket {\psi_b}$. The collection $\{p_b,\ket{\psi_b}\}$ is the ``projected ensemble''. Its average is the reduced state $\rho=\sum_bp_b\ketbra{\psi_b}{\psi_b}$, but averaging discards information. The ensemble retains how the states are spread over the Hilbert space.
For chaotic dynamics and a large measured system, this distribution of states is universal. When $\rho\propto I$, it approaches the Haar ensemble, so the states form an emergent state design~\cite{Cotler23,Choi23,Ho22,Ippoliti22,Ippoliti23}. Otherwise, it approaches the Scrooge ensemble~\cite{Jozsa94,Goldstein06,Goldstein16,Mark24,Mok26}, a deformation of Haar determined by $\rho$. Scrooge ensembles have been studied in many-body systems~\cite{McGinley25}, the Sachdev--Ye--Kitaev model~\cite{LiuZhang26}, shallow circuits~\cite{Liu26}, systems under random permutation dynamics~\cite{LiuIppolitiHo26}, and, in Gaussian versions, free fermions~\cite{Sun26,Russotto26} and bosons~\cite{LiuHuangHo24}. They have also been used to study equilibrium magic~\cite{Sarma26} and error-correcting codes~\cite{Venkatesh26}, and their properties can be predicted from few samples~\cite{Wu26}.
These works show where Scrooge appears. Here we ask which principle selects it. We show that requiring maximization to commute with local filtering determines an entropy of the ensemble, a Rényi entropy rather than Shannon's, and that Scrooge is its maximizer. 

\begin{figure}
\includegraphics[width=8.6cm]{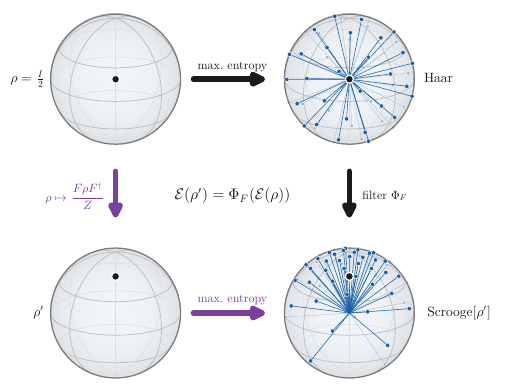}
\caption{Entropy maximization commutes with filtering, shown for a qubit. Black dots are average states, and each ensemble is drawn as points of equal weight joined to the center. Starting from $\rho=I/2$, the violet route filters the average to $\rho'\propto F\rho F^\dagger$ (left column) and then maximizes the entropy (bottom row). The black route first maximizes the entropy, which gives Haar (top row), and then filters the ensemble with $\Phi_F$ (right column). Requiring that both routes agree determines the entropy, and their common end point, $\Scr[\rho']$, satisfies Eq.~(\ref{eq:7}), written at the center.}
\label{fig:1}
\end{figure}

The canonical ensemble is the maximum-entropy ensemble at fixed mean energy~\cite{Jaynes57}. It is natural to ask which entropy plays the same role for the projected ensemble at fixed average state, a question also raised for mixed conditional states~\cite{Mok26,Yu25,Sherry26}.
Scrooge is known to minimize the accessible information~\cite{Jozsa94}, but that is not an entropy of the ensemble. At the level of the unnormalized vectors $\sqrt{p_b}\,\ket{\psi_b}$, Scrooge corresponds to a Gaussian, which maximizes their Shannon entropy at fixed second moment
\cite{Goldstein06,Goldstein16,Mark24}. The Shannon entropy of the states themselves, however, is maximized at fixed average state alone by a different ensemble~\cite{Mark24,Willson26,Shakeshaft26,Brody00,Parfionov06}.
Reference~\cite{Willson26} recovers Scrooge by also fixing the average Rényi divergence $\langle D_2(\psi\Vert\rho)\rangle$ to a prescribed value. In contrast, the principle presented here keeps the average state as the only constraint and maximizes an entropy of the ensemble of states. It derives that entropy from a consistency condition on local filters.

A ``filter'' is a local measurement on $A$ in which only one outcome is kept, like those used to concentrate entanglement~\cite{Bennett96,Verstraete01,Gisin96}. Filtering $A$ and measuring $B$ act on different systems, so they commute. But learning the outcome of a filter forces one to update the description by Bayes' rule. We ask that this update be compatible with entropy maximization (Fig.~\ref{fig:1}). Among separable entropies, only the Rényi entropy of order $\alpha_d=d/(d+1)$ satisfies this requirement. Scrooge is its unique maximizer at fixed $\rho$, and the maximum is $\ln\det(d\rho)$. When nothing but $\rho$ is known, the condition by itself selects Scrooge.

This entropy is also the maximum Shannon entropy of the unnormalized vectors that produce an ensemble, so maximizing one or the other under the same linear constraints is a single problem. The principle also gives the maximizer for mixed conditional states, for conserved charges, and for Gaussian states of free fermions.

\textit{Filters and consistency}.--- An ensemble of pure states of $A$ is a probability measure $P$ on the normalized states. We write $d\mu$ for the Haar measure, and $p=dP/d\mu$ for the density of $P$ with respect to it, when it exists.
The average state of $P$ is $\rho_P=\int dP\,\ketbra{\psi}{\psi}$, or $\rho$ when it is fixed. For full-rank $\rho$, the Scrooge ensemble has density~\cite{Jozsa94,Goldstein06}
\beq
p_{\rm Scr}(\psi)=\frac{d}{\det\rho}\,\bra{\psi}\rho^{-1}\ket{\psi}^{-(d+1)}.
\label{eq:1}
\eeq

\textbf{\Def{def:filter}} \emph{(filter):} Let $F$ be an invertible operator on $A$. The filter $\Phi_F$ takes each state $\psi$ to $F\psi/\Vert F\psi\Vert$, multiplies its weight by $\Vert F\psi\Vert^2$, which is proportional to the success probability of the filter, and then renormalizes.

Throughout, a tilde marks unnormalized objects, such as a vector $\tilde\psi\in\mathbb{C}^d$, and $\psi=\tilde\psi/\Vert\tilde\psi\Vert$ is its direction.
When $F\otimes I$ is applied to the global state and the outcome is postselected, the unnormalized vectors of the projected ensemble, $\tilde\psi_b=(I\otimes\bra{b})\ket{\Psi}=\sqrt{p_b}\,\ket{\psi_b}$, become $F\tilde\psi_b$.
 Because the filter acts on $A$ and the measurement on $B$, 
filtering before or after the measurement gives the same ensemble. Filters of this kind arise, for example, as
the Kraus operators of a subsystem's dynamics, conditioned on the classical configuration of its environment~\cite{Huang26}. Filters compose, $\Phi_{F_1}\circ\Phi_{F_2}=\Phi_{F_1F_2}$, and the average state transforms as
\beq
\rho\;\longmapsto\;\frac{F\rho F^\dagger}{\tr(F^\dagger F\rho)} .
\label{eq:2}
\eeq
A familiar case is the thermal state, which is the maximally mixed state filtered by $F=e^{-\beta H/2}$, since $e^{-\beta H}\propto F\,I\,F^\dagger$~\cite{Sugiura13}.

The density of the filtered ensemble has a closed form. For $\psi'=F\psi/\Vert F\psi\Vert$ [Appendix~\ref{app:FFEB}],
\beq
p'(\psi')=\frac{\Vert F\psi\Vert^{2(d+1)}}{Z\,|\det F|^2}\;p(\psi),\qquad Z=\tr(F^\dagger F\rho_P).
\label{eq:3}
\eeq
The exponent $d+1$ has two origins: $d$ comes from the Jacobian of the change of variables $\psi\rightarrow\psi'$, and $1$ from the Born weight.

Maximum entropy is a rule of inference. Among the ensembles compatible with what is known, one assigns the most disordered. A filter adds information, which is incorporated through Bayes' rule. If one learns that the outcome with Kraus operator $F$ occurred on $A$, the Born likelihood $\Vert F\psi\Vert^2$ forces any ensemble $P$ to be updated to $\Phi_F(P)$.
There are then two ways to describe the system after the filter. One can assign the most disordered ensemble and then update it, or update what is known and then assign. We ask that the two descriptions agree, since both use the same information.

\textbf{Consistency condition} \emph{(entropy maximization commutes with filtering):} Let $\mathcal C$ be a set of ensembles with the same average state and let $F$ be a filter. The maximum-entropy ensemble of $\Phi_F(\mathcal C)$ is $\Phi_F$ of the maximum-entropy ensemble of $\mathcal C$.

The set $\mathcal C$ represents what is known: the average state (which is the single-copy constraint), and any additional constraints. Taking $\mathcal C=\{P_1,P_2\}$ shows that the condition is equivalent to the following: \textit{a filter does not change which of two ensembles with the same average is more disordered}.

On ensembles, a filter can be undone, since $\Phi_{F^{-1}}$ inverts $\Phi_{F}$. The ensembles before and after are in one-to-one correspondence, and no information about which ensemble is present is created or destroyed.
Moreover, projected ensembles satisfy the condition exactly.
For Haar-random global states with a given reduced state, filtering takes the projected ensemble at $\rho$ to the one at $\rho'$ \smcite{sm:CC}.

The condition is not a symmetry. A filter that is not proportional to a unitary does not preserve the Haar measure and in general changes the average state, so one cannot ask for an invariant entropy. One can only ask that the change of the entropy be the same for all the ensembles being compared. We take the consistency condition as a postulate and derive its consequences.

\textit{Entropy and covariance}.--- We measure the disorder of an ensemble with the Rényi
entropies relative to Haar~\cite{Mark24}, $S_\alpha(P)=-D_\alpha(P\Vert\Haar)$, where
\beq
D_{\alpha}(P\Vert P_0)=\frac{1}{\alpha-1}\ln\int dP_0 \left(\frac{dP}{dP_0}\right)^\alpha,\qquad0<\alpha<1,
\label{eq:4}
\eeq
with $\frac{dP}{dP_0}$ the density of the part of $P$ that is absolutely continuous with respect to $P_0$ (a singular part does not contribute). The entropies $S_\alpha$ are nonpositive, vanish only for Haar [Appendix~\ref{app:FFEB}], and equal $-\infty$ for discrete ensembles. The limit $\alpha\to1$ gives the Shannon entropy, $S\equiv S_1$, and the Kullback--Leibler divergence $D_1\equiv D_{\rm KL}$. For a reference $P_0$ other than Haar we write $S_\alpha(P\vert P_0)=-D_\alpha(P\Vert P_0)$, so that $S_\alpha(P) =S_\alpha(P\vert \text{Haar})$.\medskip

\textbf{\Lem{lem:cov}} \emph{(covariance law).} Let $\alpha_d=d/(d+1)$. Given any ensemble $P$, with
average state $\rho_P$, and any filter $F$,
\beq
\begin{aligned}S_{\alpha_d}(\Phi_F(P))={}&S_{\alpha_d}(P)\\&-d\ln\tr(F^\dagger F\rho_P)+2\ln|\det F|.\end{aligned}
\label{eq:5}
\eeq

\textit{Proof.} Let $p'$ and $Z$ be as in Eq.~(\ref{eq:3}). Using Eq.~(\ref{eq:3}) and the change of variables $d\mu(\psi')=|\det F|^2\Vert F\psi\Vert^{-2d} d\mu(\psi)$ [Appendix~\ref{app:FFEB}], we find $\int d\mu\,p'^{\,\alpha}=Z^{-\alpha} |\det F|^{2(1-\alpha)} \int
d\mu\,p^\alpha\,\Vert F\psi\Vert^{2(d+1)\alpha-2d}$. The power of $\Vert F\psi\Vert$
vanishes if and only if $\alpha=\alpha_d$. Taking the logarithm and multiplying by
$1/(1-\alpha_d)=d+1$ then gives the result. $\square$

\begin{figure}
\includegraphics[width=8.6cm]{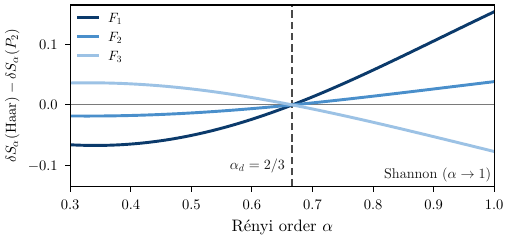}
\caption{A filter shifts the entropy of order $\alpha_d=d/(d+1)$ by the same amount for all ensembles with the same average. For a qubit, we compare Haar with $P_2$, of density $1+0.6(3z^2-1)$ relative to Haar ($z$ the Bloch coordinate), both with average $I/2$. The filters are $F_j=R_j\,{\rm diag}(1,e^{-1})R_j^\dagger$, where $R_j$ tilts the axis from $z$ by $0$, $\pi/4$ and $\pi/2$ ($j=1,2,3$). The difference $\delta S_\alpha(\Haar)-\delta S_\alpha(P_2)$, with $\delta S_\alpha(P)=S_\alpha(\Phi_{F_j}(P))-S_\alpha(P)$, vanishes at $\alpha_d=2/3$ for all three filters (Lemma~\ref{lem:cov}) and at no other order.}
\label{fig:2}
\end{figure}

A filter shifts $S_{\alpha_d}$ by the same amount for all ensembles with the same average, whereas for any other order the shift depends on the ensemble (Fig.~\ref{fig:2}). The ranking of these ensembles is therefore unchanged, and $S_{\alpha_d}$ satisfies the consistency condition. It is the only separable entropy that does, where an entropy is separable if it
has the form $\mathcal H_\varphi(P)=\int d\mu\,\varphi(p)$. The Shannon entropy corresponds to $\varphi(s)=-s\ln s$, and the Rényi and Tsallis entropies of order $\alpha$ are monotone functions of $\mathcal H_\varphi$ with $\varphi(s)=s^\alpha$. We only ask that $\varphi$ be continuously differentiable. \medskip

\textbf{\Thm{thm:order}.} Let $d\ge2$, and let $\varphi$ be continuously differentiable on $(0,\infty)$. If every filter preserves the ranking by $\mathcal H_\varphi$ of ensembles with the same average state, that is, if $\mathcal H_\varphi(P_1)>\mathcal H_\varphi(P_2)$ implies $\mathcal H_\varphi(\Phi_F(P_1))>\mathcal H_\varphi(\Phi_F(P_2))$, then $\varphi(s)=a_0+a_1\,s+a_2\,s^{d/(d+1)}$. Hence $\mathcal H_\varphi$ is constant ($\varphi$ affine) or a monotone function of the Rényi entropy of order $d/(d+1)$, increasing when $\varphi$ is concave ($a_2>0$). In particular, neither the Shannon entropy nor the Rényi or Tsallis entropies of any other order satisfy the hypothesis.

Indeed, since $\int d\mu\,p=1$, the terms $a_0$ and $a_1$ are constants and $\mathcal H_\varphi=a_0+a_1+a_2\,e^{S_{\alpha_d}/(d+1)}$. The Rényi entropy of order $d/(d+1)$ is therefore the only separable entropy for which maximizing and filtering commute. The proof is given in \smcite{sm:TI}. The order reflects the Born rule. By Eq.~(\ref{eq:3}), a filter multiplies the density by a state-dependent factor with exponent $d+1$. For that factor to cancel, $\varphi$ must be a power, and among the powers it cancels only for $\alpha=d/(d+1)$ (Fig.~\ref{fig:2}). Other Born exponents would give other orders, but then the filtered average would not be a function of $\rho$ alone \smcite{sm:GW}.

\textit{The Scrooge ensemble}.--- The consistency condition determined the entropy with no reference to Scrooge. Yet maximizing this entropy at a given average state $\rho$ gives the Scrooge ensemble.\medskip

\textbf{\Thm{thm:max}.} Among all ensembles with full-rank average state $\rho$, $S_{\alpha_d}$ attains its maximum, $\ln\det(d\rho)$, only at $\Scr[\rho]$.

\textit{Proof.} Let $P$ have average $\rho$ and $F=\rho^{-1/2}$. By Eq.~(\ref{eq:2}), $\Phi_F(P)$ has average $I/d$, and by Eq.~(\ref{eq:5}), with $\tr(F^\dagger F\rho)=d$ and $|\det F|^2=1/\det\rho$,
\beq
S_{\alpha_d}(P)=S_{\alpha_d}(\Phi_F(P))+\ln\det(d\rho)\le\ln\det(d\rho),
\label{eq:6}
\eeq
because $S_{\alpha_d}\le0$. Equality holds if and only if $\Phi_F(P)$ is Haar, that is, $P=\Phi_{\sqrt\rho}(\Haar)$. By Eq.~(\ref{eq:3}) with $p=1$ and $F=\sqrt\rho$, that ensemble has the density of Eq.~(\ref{eq:1}). $\square$

The filter $\rho^{-1/2}$ thus moves every ensemble with average $\rho$ to average $I/d$, where the maximizer is Haar, and shifts all their entropies by the same amount, $-\ln\det(d\rho)$. The inverse filter then takes Haar to Scrooge. The argument needs no regularity, so it also covers ensembles with discrete or singular parts. The value $\ln\det(d\rho)$ is also the maximum Shannon entropy of the unnormalized vectors, relative to a Gaussian reference~\cite{Mark24}, a coincidence that Theorem~\ref{thm:SR} explains. Finally, the distance to the maximum, $\ln\det(d\rho)- S_{\alpha_d}(P)$, is invariant under filters, because by Lemma~\ref{lem:cov} both terms shift equally. 

If $\rho$ is all that is known, a rule assigns an ensemble $\mathcal E(\rho)$ to every average state $\rho$. The condition then says that assigning and filtering should commute (Fig.~\ref{fig:1}),
\beq
\mathcal E(\rho')=\Phi_F\big(\mathcal E(\rho)\big),\qquad\rho'=\frac{F\rho F^\dagger}{\tr(F^\dagger F\rho)}.
\label{eq:7}
\eeq
We call a rule $\mathcal E$ that satisfies Eq.~(\ref{eq:7}) covariant. Covariance is enough to determine the rule, with no entropy involved.

\textbf{\Obs{obs:rule}.} Let a rule $\mathcal E$ assign an ensemble of pure states to every full-rank $\rho$. It satisfies Eq.~(\ref{eq:7}) for every invertible filter $F$ if and only if $\mathcal E(\rho)=\Scr[\rho]$.

\textit{Proof.} Unitaries are filters. With $\rho=I/d$ and $F=U$, covariance says that $\mathcal E(I/d)$ is invariant under every unitary, so it is Haar. Every full-rank $\rho$ is $I/d$ filtered by $\sqrt\rho$, and hence $\mathcal E(\rho)=\Phi_{\sqrt\rho}(\Haar)=\Scr[\rho]$. Conversely, this rule is covariant. By polar decomposition, $F\sqrt\rho=\sqrt{Z\rho'}\,U$ with $U$ unitary, and $\Phi_U$ leaves Haar fixed, so $\Phi_F\circ \Phi_{\sqrt\rho}(\Haar)=\Phi_{\sqrt{\rho'}}(\Haar)$. $\square$

Scrooge was defined as $\Phi_{\sqrt\rho}(\Haar)$~\cite{Jozsa94}, that is, as Haar-random states filtered by the square root of the average state. 
Reference~\cite{Goldstein06} conjectured, and proved for continuous families, that Scrooge is the only unitarily covariant and hereditary family of ensembles. Every filter is also known to take Scrooge to Scrooge~\cite{Tumulka26}. Observation~\ref{obs:rule} gives the converse, that Scrooge is the only covariant rule. The consistency condition thus fixes the entropy (Theorem~\ref{thm:order}). It also fixes the ensemble when only $\rho$ is given (Observation~\ref{obs:rule}), and the two agree (Theorem~\ref{thm:max}). Maximizing any other $S_\alpha$, Shannon included, does not commute with filters \smcite{sm:OI}.

\textit{From Shannon to Rényi}.--- The measurement produces the unnormalized vectors $\tilde\psi_b=\sqrt{p_b}\,\ket{\psi_b}$. On them a filter is simply the linear map $\tilde\psi\mapsto F\tilde\psi$, with no reweighting, and the Born weight appears only upon normalizing. 
The Shannon entropy of these vectors at fixed second moment is maximized by a Gaussian, and its maximum relative to a Gaussian reference is $\ln \det (d\rho)$~\cite{Mark24}. Theorem~\ref{thm:SR} shows that the Rényi entropy of the states is the maximum Shannon entropy of the vectors that produce them. We state it for a general reference, because it then covers pure and mixed states at once.

\textbf{\Def{def:produce}.} Let $P_0$ be a reference probability on a space $X$ (for pure states, Haar) and $P$ another one, with density $p=dP/dP_0$. A probability $\tilde P$ on $X\times(0,\infty)$ \textit{produces} $P$ with Born exponent $\nu>0$ if $\int_{\mathcal A\times (0,\infty)} d\tilde P\,w^{\nu}=P(\mathcal A)$ for every measurable $\mathcal A\subset X$, and we then write $\tilde P\to P$. 

For a fixed $c>0$, we consider the reference $\tilde P_0=P_0\otimes G$ on $X\times (0,\infty)$, with $\frac{dG}{dw}=\frac{\nu\,c^{c}}{\Gamma(c)}\,w^{\nu c-1}e^{-c\,w^{\nu}}$, which is normalized and has $\mathbb E_{\tilde P_0}[w^{\nu}]=1$, so that $\tilde P_0\to P_0$. Let $\tilde p$ be the density $d\tilde P/d\tilde P_0$. The Shannon entropy of $\tilde P$ relative to $\tilde P_0$ is $S(\tilde P|\tilde P_0)=-\int d\tilde P\ln(\tilde p)=-D_{\rm KL}(\tilde P\Vert\tilde P_0)$.\medskip

\textbf{\Thm{thm:SR}} \emph{(Shannon--Rényi connection).} Consider ensembles $P$ and $\tilde P$ with densities with respect to $P_0$ and $\tilde P_0$, respectively.

(a) For every $P$, the minimum Kullback--Leibler divergence to $\tilde P_0$ among the $\tilde P$ that produce it is its Rényi divergence of order $\alpha_c=c/(c+1)$ to $P_0$:
\beq
\min_{\tilde P\to P}D_{\rm KL}(\tilde P\Vert\tilde P_0)=D_{\alpha_c}(P\Vert P_0).
\label{eq:8}
\eeq
The minimum is attained at a unique $\tilde P_*$, whose density with respect to $\tilde P_0$ is $\tilde p_*=\kappa\,e^{-\lambda(x)w^{\nu}}$, with $\kappa=e^{-S_{\alpha_c}(P|P_0)}$ and $1+\lambda(x)/c=(p(x)/\kappa)^{-1/(c+1)}$.

(b) Let $h_1,\dots,h_l$ be real functions on $X$ with associated constraint values $\bar h_1,\dots,\bar h_l$. Then
\beq
\max_{\int d\tilde P\,w^{\nu}h_i=\bar h_i}S(\tilde P|\tilde P_0)=\max_{\int dP\,h_i=\bar h_i}S_{\alpha_c}(P|P_0),
\nonumber
\eeq
and the maximum on the left is attained at the $\tilde P_*$ of (a) built from the maximizer on the right.

Part (b) says that the two maximum-entropy problems are equivalent. The unnormalized ensemble of maximum Shannon entropy under the linear constraints $\int d\tilde P\,w^{\nu}h_i=\bar h_i$ corresponds to the normalized ensemble of maximum Rényi entropy of order $\alpha_c$ under the same constraints, $\int dP\,h_i=\bar h_i$. It follows from (a) because the constraint $\int d\tilde P\,w^{\nu}h_i=\int dP\,h_i$ depends on $\tilde P$ only through $P$, and $\max_{\tilde P}=\max_P\max_{\tilde P\to P}$. The proof is in Appendix~\ref{app:TIII}. Since $S(\tilde P|\tilde P_0)=-D_{\rm KL}(\tilde P\Vert\tilde P_0)$ and $S_\alpha(P|P_0)=-D_\alpha(P\Vert P_0)$ [Eq.~(\ref{eq:4})], Eq.~(\ref{eq:8}) reads $\max_{\tilde P\to P}S(\tilde P|\tilde P_0)=S_{\alpha_c}(P|P_0)$.

For pure states, let $X$ be the space of rays, $P_0=$ Haar, $\nu=2$ and $c=d$. Then $\tilde P$ is an unnormalized ensemble, that is, a probability on vectors $\tilde\psi\in\mathbb C^d$ with $\mathbb E_{\tilde P}[\Vert\tilde\psi\Vert^2]=1$, written in the coordinates $\tilde\psi=w\psi$. Producing $P$ means normalizing each vector and weighting it by $\Vert\tilde\psi\Vert^2$, the Born weight. The reference $\tilde P_0$ is the Gaussian $\gamma=\mathcal{CN}(0,I/d)$, whose radial density is $dG/dw\propto w^{2d-1}e^{-dw^2}$, which is why $\nu=2$ and $c=d$. With $S(\tilde P|\gamma)=-D_{\rm KL}(\tilde P\Vert\gamma)$~\cite{Mark24} and $\alpha_d=d/(d+1)$, Eq.~(\ref{eq:8}) says that the maximum Shannon entropy of the unnormalized vectors that produce $P$ is $S_{\alpha_d}(P)$, and that the maximizer is Gaussian along each ray.

The theorem also applies to ensembles of mixed states, which we treat below. There, an unmeasured register of dimension $K$ induces a reference measure $P_K$ on the states of $A$. Let $X$ be those states, $P_0=P_K$, $\nu=1$ and $c=dK$. Then $w\sigma$ is a positive matrix with trace $w$, and $\tilde P_0$ is the law of the Wishart matrix with average $I/d$, namely $Y\,Y^\dagger$ with $Y$ a $d\times K$ Gaussian matrix of variance $1/(dK)$.
The order is $\alpha_{dK}=dK/(dK+1)$, the same as in Theorem~\ref{thm:mixed} (see below). By part (b) with $h=\sigma$, maximizing the Shannon entropy at fixed first moment $\mathbb E_{\tilde P}[w\sigma]=\rho$ is equivalent to maximizing $S_{\alpha_{dK}}$ at fixed average $\rho$. The maximizer of the former is a correlated Wishart ensemble~\cite{Vinayak10,Burda05}. In both cases the order is $c/(c+1)$, with $c$ read off from the radial density of the reference, $w^{\nu c-1}$.

\textit{Mixed states}.--- If a register $C$ of dimension $K$ is left unmeasured, the conditional states $\sigma_b$ of $A$ are mixed. The natural reference is then the induced measure $P_K$~\cite{Sommers04,Zyczkowski11,Lubkin78,Hall98,Zyczkowski01}, the distribution of $\tr_C \ketbra{\chi}{\chi}$ with $\ket\chi$ Haar on $A\otimes C$. A filter on $A$ acts as $F\otimes I$, and the success probability, $\tr(F^\dagger F\sigma)$, depends only on $\sigma$. Every ensemble of mixed states of $A$ can be ``purified'': each state $\sigma$ is replaced by its purifications in $A\otimes C$, spread uniformly over them. Two facts carry Lemma~\ref{lem:cov} over to the mixed case. First, the entropy does not change. The entropy of a mixed ensemble relative to $P_K$ equals that of the purified ensemble relative to Haar on $A\otimes C$. The purifications add degrees of freedom, but on them the ensemble and Haar coincide, and the only thing that distinguishes the two is the choice of $\sigma$. Second, filtering and purifying commute. The operator $F\otimes I$ lets all purifications of $\sigma$ succeed with the same probability, $\tr(F^\dagger F\sigma)$, and takes them to purifications of $F\sigma F^\dagger/\tr(F^\dagger F\sigma)$. Lemma~\ref{lem:cov} in dimension $dK$ then gives the covariance law for mixed states, with $d\to dK$ and $|\det F|\to|\det F|^K$, since $\det(F\otimes I)=(\det F)^K$. As before, the proof of Theorem~\ref{thm:order} goes through with $d\to dK$ [Appendix~\ref{app:MSCC}], so no other separable entropy relative to $P_K$ has this property and the order is $\alpha_{dK}=dK/(dK+1)$. Theorem~\ref{thm:max} applied to the purified ensemble then gives the following. \medskip

\textbf{\Thm{thm:mixed}.} Among the ensembles of mixed states of $A$ with full-rank average $\rho$, $S_{\alpha_{dK}}(P|P_K)$ attains its maximum, $K\ln\det(d\rho)$, only at
\beq
\Scr^{(K)}[\rho]=\tr_C\,\Scr[\rho\otimes I/K].
\label{eq:11}
\eeq

Unlike in the pure case, at $\rho=I/d$ unitary symmetry does not determine the answer, because an ensemble with Haar-random eigenvectors is invariant whatever the distribution of its spectrum. Covariance carries the answer at $I/d$ to every $\rho$, but that answer, $P_K$, is a physical input that encodes how many dimensions are left unmeasured. Equation~(\ref{eq:11}) is exactly the random-POVM ensemble of Ref.~\cite{Shakeshaft26}, whose density relative to $P_K$ is proportional to $[\tr(\rho^{-1}\sigma)]^{-(dK+1)}$. It is also a correlated Wishart ensemble~\cite{Vinayak10,Burda05}. If $W$ is Wishart with $K$ degrees of freedom and mean $\rho$, the states $\sigma=W/\tr\,W$, weighted by the Born weight $\tr\,W$, have that density. The maximizer of the Shannon entropy relative to the same reference $P_K$ is a different ensemble, with the same large-dimension limit. For $\rho=I/d$, Eq.~(\ref{eq:11}) reduces to $P_K$, the generalized Hilbert--Schmidt ensemble of Ref.~\cite{Yu25}. Theorem~\ref{thm:mixed} gives a maximum-entropy principle for both the random-POVM and the generalized Hilbert--Schmidt ensembles. The ensembles of error channels~\cite{Yu26} and of mixed initial states~\cite{Sherry26} have the same form, $\tr_C\,\Scr[\Sigma]$, with structure on the hidden side.

For a conserved charge that the measurement reveals, covariance under the filters that respect it selects a Scrooge ensemble in each charge sector~\cite{Mark24,Chang25}, and the entropy maximized in a sector of dimension $d_Q$ has order $d_Q/(d_Q+1)$ \smcite{sm:CQ}.

\textit{Free fermions}.--- Our results extend to other sets of states and filters, such as the Gaussian states of $N\geq 2$ fermionic modes, with Majoranas $\gamma_1,\dots,\gamma_{2N}$ and $d=2^N$. The pure Gaussian states of parity $\pm$ form manifolds $\mathcal M_\pm$, of complex dimension $N(N-1)/2$, on which the Gaussian unitaries act transitively~\cite{SuraceTagliacozzo22,HacklBianchi21}, and $\mu_G=\frac12(\mu_++\mu_-)$, with $\mu_\pm$ invariant, plays the role of Haar. The filters are the Gaussian operators, products of $\zeta\,e^{\hat X}$, with $\zeta\in\mathbb C^*$, $\hat X=\frac12\sum_{ab}X_{ab}\gamma_a\gamma_b$ and $X$ complex antisymmetric. They map Gaussian states to Gaussian states, and include $\sqrt\rho$ for every full-rank Gaussian $\rho$. Projected ensembles of free fermions measured in a Gaussian basis consist of Gaussian states. They approach $\mu_G$~\cite{Bejan25} or its generalizations~\cite{Lucas23}, and they approach the fermionic Gaussian Scrooge ensemble $\Phi_{\sqrt\rho}(\mu_G)$ at fixed $\rho$~\cite{Sun26,Russotto26}. 
As for pure states, the filtered density has a closed form. For a Gaussian filter $F$ and $\psi\in\mathcal M_\pm$,
\beq
p'(\psi')=\frac{\Vert F\psi\Vert^{2(r+1)}}{Z\,|\det F|^{2r/d}}\;p(\psi),\qquad r=2(N-1),
\label{eq:12}
\eeq
with densities relative to $\mu_G$ and $Z$ as in Eq.~(\ref{eq:3}). \medskip

\textbf{\Thm{thm:fermions}.} For ensembles of Gaussian states and Gaussian filters, Lemma~\ref{lem:cov} and Theorems~\ref{thm:order} and~\ref{thm:max} hold with Haar$\,\to\mu_G$, $d\to r$ and $|\det F|\to|\det F|^{r/d}$. In particular, at fixed full-rank Gaussian $\rho$, $S_{\alpha_r}(P|\mu_G)$, with $\alpha_r=r/(r+1)$, attains its maximum, $\frac rd\ln\det(d\rho)$, only at the fermionic Gaussian Scrooge ensemble, which is also the only covariant rule.

The proofs are in Appendix~\ref{app:GS} and \smcite{sm:GS}, where we also show that Gaussian projected ensembles satisfy the consistency condition exactly. The restriction to Gaussian states, which the measurement provides, is essential. Without it the covariant rule would not be unique, because Scrooge is also covariant under Gaussian filters.
As the Gaussian states form a much smaller manifold, $r$ replaces $d$ in the order and grows only linearly with the number of modes.

\textit{Discussion.}--- Scrooge is a maximum-entropy ensemble in the usual sense, that is, the unique maximizer of an entropy of the ensemble of states at fixed average. That entropy is the Rényi entropy of order $d/(d+1)$ relative to Haar. It is determined by a physical consistency condition, that maximizing and filtering commute, rather than chosen to reproduce Scrooge. Its order encodes the Born rule, with $d$ coming from the Jacobian and 1 from the weight. The same condition gives the maximizers for mixed states (given the reference $P_K$), for conserved charges, and for free fermions.

The principle is one of inference. It determines which ensemble to assign when only the average state $\rho$ is known, and it makes no assumption about the underlying dynamics. This entropy is unique among separable entropies, whereas the covariance of Scrooge extends to noninvertible filters, with the order adjusting to the rank of the filter. Scrooge thus maximizes entropy at fixed average state, as the canonical ensemble does at fixed mean energy.

\textit{Acknowledgments.}---This work was partially supported by CONICET, UBACyT (20020220300049BA and 20020250300103BA).
The authors used Claude (Anthropic) as an assistant for literature search, analytic and numerical checks, figure scripts and drafting. The authors verified the results and are responsible for the content.

\section*{End Matter}

\appsec{Filter formula and entropy bound}{app:FFEB}

\textit{The change of variables.} Let $T(\psi)=F\psi/\Vert F\psi\Vert$. We use a standard fact about Gaussian integrals \smcite{sm:GT}. If $\tilde\psi\in\mathbb C^d$ is a Gaussian vector, $\tilde\psi\sim\mathcal{CN}(0,\Sigma)$, then its direction $\psi=\tilde\psi/\Vert\tilde\psi\Vert$ has density $(\det\Sigma)^{-1}\bra{\psi}\Sigma^{-1}\ket{\psi}^{-d}$ relative to Haar. In particular, if $\tilde\psi\sim\mathcal{CN}(0,I)$, its direction is Haar and $F\tilde\psi\sim\mathcal{CN}(0,FF^\dagger)$. The image of Haar under $T$ has density $|\det F|^{-2}\bra{\psi'}(FF^\dagger)^{-1}\ket{\psi'}^{-d}$. Since $\bra{\psi'}(FF^\dagger)^{-1}\ket{\psi'}=\Vert F^{-1}\psi'\Vert^2=\Vert F\psi\Vert^{-2}$ for $\psi'=T(\psi)$, this is 
\beq
d\mu(\psi')=|\det F|^{2}\,\Vert F\psi\Vert^{-2d}\,d\mu(\psi).\tag{\ref{app:FFEB}1}
\label{eq:FFEB1}
\eeq

\textit{Eq.~(\ref{eq:3}).} The ensemble $\Phi_F(P)$ is the image under $T$ of the measure $\Vert F\psi\Vert^2P(d\psi)/Z$, with $Z=\int dP\Vert F\psi\Vert^2=\tr(F^\dagger F\rho_P)$. If $P=p\,\mu+P_s$ with $P_s$ singular, the image of $P_s$ is again singular, because $T$ is a diffeomorphism. The part with a density has density $\Vert F\psi\Vert^2p(\psi)/Z$ divided by the factor in Eq.~(\ref{eq:FFEB1}), which is Eq.~(\ref{eq:3}).

\textit{The bound $S_\alpha\le0$.} This is the positivity of the Rényi divergence \cite[Theorem 8]{vanErven14}. Explicitly, for every $P$, Jensen's inequality gives $\int d\mu\,p^\alpha\le(\int d\mu\,p)^\alpha\le1$, with equality if and only if $p\equiv1$ and there is no singular part. Thus $S_{\alpha}\le0$, with equality only for Haar, as used in the proof of Theorem~\ref{thm:max}.

\appsec{Proof of Theorem~\ref{thm:SR}}{app:TIII}

Theorem~\ref{thm:SR} is equivalent to the statement that $S(\tilde P|\tilde P_0)\le S_{\alpha_c}(P|P_0)$ for every $\tilde P\to P$, with equality only at $\tilde P_*$. Here $S(\tilde P|\tilde P_0)=-D_{\rm KL}(\tilde P\Vert\tilde P_0)$ and $S_{\alpha_c}(P|P_0)=-D_{\alpha_c}(P\Vert P_0)$. We write $\alpha_c=c/(c+1)$, so that $1-\alpha_c=1/(c+1)$.

The condition that $\tilde P$ produces $P$ fixes, for each $x$, the integral $\int dG\,w^{\nu}\,\tilde p(x,w)=p(x)$. Introducing a multiplier $\lambda(x)$ for each point and one for the normalization, and maximizing $-\int d \tilde P\ln(\tilde p)$, we obtain
\beq
\tilde p_*(x,w)=\kappa \, e^{-\lambda(x)w^{\nu}}.\tag{\ref{app:TIII}1}
\label{eq:TIII1}
\eeq
At each $x$, $\tilde P_*$ is the reference deformed by an exponential in $w^{\nu}$, which is a Gaussian when $\nu=2$.

Two integrals in $w$ are needed, $J_0(s)=\int dG\,e^{-sw^{\nu}}=(1+s/c)^{-c}$ and $J_1(s)=\int dG\,w^{\nu}\,e^{-sw^{\nu}}=(1+s/c)^{-(c+1)}$. The normalization and the constraint read $1=\kappa\int dP_0\,J_0(\lambda)$ and $p=\kappa\,J_1(\lambda)$. The second gives $1+\lambda/c=(p/\kappa)^{-1/(c+1)}$. Substituting in the first, $1=\kappa^{1-\alpha_c}\int dP_0\,p^{\alpha_c}$, that is, $\kappa=e^{-S_{\alpha_c}(P|P_0)}$.

From Eq.~(\ref{eq:TIII1}), $S(\tilde P_*|\tilde P_0)=-\ln\kappa+\int dP_0\,p\,\lambda$. The last term vanishes by the same relation between $p$ and $\lambda$, because $\int dP_0\,p\,(1+\lambda/c)=\kappa^{1/(c+1)}\int dP_0\,p^{\alpha_c}=1$. For any $\tilde P\to P$, the quantity $\int d \tilde P \ln(\tilde p_*) =\ln\kappa-\int dP_0\,p\,\lambda$ takes the same value, because it depends only on the Born weight at each $x$, which is fixed. Hence $S(\tilde P|\tilde P_0)=S_{\alpha_c}(P|P_0)-D_{\rm KL}(\tilde P\Vert\tilde P_*)$, and Gibbs' inequality gives the maximum, attained only at $\tilde P_*$. For pure states, where $\nu=2$ and $c=d$, $\tilde P_*$ is Gaussian along each ray, with its own width. $\square$

\appsec{Mixed states}{app:MSCC}

\textit{Purified ensemble.} We write $\chi$ for the pure states of $A\otimes C$ and $\sigma(\chi)=\tr_C\ketbra{\chi}{\chi}$ for their reduced state on $A$. An ensemble $P$ of mixed states of $A$ is \textit{purified} to an ensemble of pure states of $A\otimes C$ by spreading each $\sigma$ uniformly over its purifications. The purified ensemble has three properties. (a) \textit{It has the same entropy.} If $p$ is the density of $P$ relative to $P_K$, the density of the purified ensemble relative to Haar is $p(\sigma(\chi))$. Since $P_K$ is the image of Haar under $\chi\mapsto\sigma(\chi)$, it follows that $\int d\mu_{AC}(p\circ\sigma)^\alpha=\int dP_K p^\alpha$. (b) \textit{Its average is $\rho \otimes I/K$.} The purified ensemble is invariant under the unitaries on $C$, so its average is of the form $O\otimes I$, and its partial trace is $\rho$. (c) \emph{It determines $P$.} The ensemble $P$ is recovered from the purified one by taking $\tr_C$ of each state.

\textit{Proof of Theorem~\ref{thm:mixed}.} This is Theorem~\ref{thm:max} in dimension $dK$, applied to the purified ensemble. By (a) and (b), the entropy of $P$ is at most $\ln\det(dK\,\rho\otimes I/K)=K\ln\det(d\rho)$, with equality only if the purified ensemble is $\Scr[\rho\otimes I/K]$. That ensemble is invariant under the unitaries on $C$, because its density, Eq.~(\ref{eq:1}), depends on $\chi$ only through $\sigma(\chi)$. It is therefore the purification of its partial trace, and by (c) $P$ is the ensemble of Eq.~(\ref{eq:11}). $\square$ 

\textit{Filters.} For every purification $\chi$ of $\sigma$, $\Vert(F\otimes I)\chi\Vert^2=\tr(F^\dagger F\sigma)$, and the reduced state of the normalized $(F\otimes I)\chi$ is $F\sigma F^\dagger/\tr(F^\dagger F\sigma)$, so filtering and purifying commute. Lemma~\ref{lem:cov} in dimension $dK$, with $|\det(F\otimes I)|^2=|\det F|^{2K}$, then gives the covariance law for mixed states, relative to $P_K$ and only for the order $\alpha_{dK}$: 
$S_{\alpha_{dK}}(\Phi_F(P)|P_K)=S_{\alpha_{dK}}(P|P_K)-dK\ln\tr(F^\dagger F\rho)+2K\ln|\det F|$. If $K<d$, $P_K$ is supported on states of rank at most $K$, and the part of an ensemble on states of higher rank is singular with respect to $P_K$ and only lowers the entropy.

\textit{Uniqueness of the order.} The proof of Theorem~\ref{thm:order} \smcite{sm:TI} carries over to mixed states with $\mu\to P_K$ and $d\to dK$. It rests on three facts, and each one has a mixed-state counterpart. (i) \emph{A variable with a positive density.} In place of $|\langle0|\psi\rangle|^2$ we use $x=\langle0|\sigma|0\rangle$. Under $P_K$ its density is proportional to $x^{K-1}(1-x)^{(d-1)K-1}$, which is continuous and positive on $(0,1)$. (ii) \emph{A success probability linear in $x$.} For $F={\rm diag}(1,f,\dots,f)$ with $0<f<1$, $t=\tr(F^\dagger F\sigma)=x+f^2(1-x)$. (iii) \emph{The change of variables.} A purification of $\sigma$ is a pure state in dimension $dK$, filtered by $F\otimes I$. Since $P_K$ is the image of the Haar measure under the partial trace, $\int dP_K\,\varphi(p)$ takes the same value on the purified ensemble for every $\varphi$, and Appendix~\ref{app:FFEB} applies in dimension $dK$, where the filter multiplies the density by $t^{dK+1}/(Z|\det F|^{2K})$ and the measure by $|\det F|^{2K}t^{-dK}$. The argument of \smcite{sm:TI} then gives the same equation with $d\to dK$, $\varphi'(c_F\,t^{dK+1})\,t=\kappa_0+\kappa_1t$, whose solutions are $\varphi(s)=a_0+a_1s+a_2s^{\alpha_{dK}}$. As $f\to0$ the range of $s$ on which this holds grows to $(0,\infty)$. This is valid for every $d\ge2$ and $K\ge1$.

\appsec{Gaussian states}{app:GS}

\textit{Three facts.} Let $N\ge2$, $\hat N$ be the number operator, and $\Pi_\pm=\frac12[I\pm(-1)^{\hat N}]$ the projectors onto the parity sectors $A_\pm$. (i)~\emph{Polar decomposition.} Since $(\gamma_a\gamma_b)^\dagger=-\gamma_a\gamma_b$ for $a\ne b$, the adjoint of $\zeta e^{\hat X}$ is $\bar\zeta e^{-\frac12\sum_{ab}\bar X_{ab}\gamma_a\gamma_b}$, so the group of Gaussian filters, $\mathbb C^*\cdot{\rm Spin}(2N,\mathbb C)$, 
is closed under $\dagger$. Every filter is therefore $F=\zeta e^{\hat X}U$, with $\zeta>0$, $\hat X$ Hermitian and $U$ a Gaussian unitary~\cite{Mostow55}. The positive filters $\zeta e^{\hat X}$ are unnormalized Gaussian states, so $F\rho F^\dagger/Z$ is Gaussian if $\rho$ is. (ii)~\emph{Equal determinants}, $\det_{A_+}F=\det_{A_-}F$. Since determinants are multiplicative and $\zeta$ contributes $\zeta^{d/2}$ to each sector, it suffices to take $F=e^{\hat X}$, for which $\det_{A_\pm}F=e^{\tr(\Pi_\pm\hat X)}$. Up to a phase, $(-1)^{\hat N}$ is the product of all $2N$ Majoranas, so $\gamma_a\gamma_b$ and $(-1)^{\hat N}\gamma_a\gamma_b$, for $a\ne b$, are products of an even number $k\ge2$ of distinct Majoranas, traceless by anticommutation and cyclicity. Hence $\tr(\Pi_\pm\hat X)=0$. (iii)~\emph{Average of $\mu_G$}. The average of $\mu_\pm$ is invariant under the Gaussian unitaries, so it commutes with every $\gamma_a\gamma_b$. Their products give all the products of an even number of Majoranas, which span the parity-preserving operators. The average of $\mu_\pm$ is therefore $\Pi_\pm/(d/2)$, and that of $\mu_G$ is $I/d$.

\textit{Covariance law (Lemma~\ref{lem:cov}).} Write $P=\sum_\pm q_\pm P^{(\pm)}$, with $P^{(\pm)}$ on $\mathcal M_\pm$, $\rho_\pm=\Pi_\pm\rho_P\Pi_\pm$, $q_\pm=\tr\rho_\pm$, and $Z_\pm=\tr(F^\dagger F\rho_\pm)/q_\pm$. Since $\mu_G=\frac12(\mu_++\mu_-)$,
\beq
e^{(1-\alpha)S_\alpha(P|\mu_G)}=\sum_\pm\tfrac12(2q_\pm)^\alpha\,e^{(1-\alpha)S_\alpha(P^{(\pm)}|\mu_\pm)}.\tag{\ref{app:GS}1}
\label{eq:GS1}
\eeq
The filter acts within each sector. It changes the weights to $q_\pm Z_\pm/Z$, with $Z=\tr(F^\dagger F\rho_P)$, and, by the proof of Lemma~\ref{lem:cov} with Eq.~(\ref{eq:12}) in place of Eq.~(\ref{eq:3}) and the change of variables of \smcite{sm:GS} in place of Eq.~(\ref{eq:FFEB1}), it shifts $S_{\alpha_r}(P^{(\pm)}|\mu_\pm)$ by $-r\ln Z_\pm+\frac{2r}{d}\ln|\det F|$, the same in both sectors by (ii). Since $(1-\alpha_r)r =\alpha_r$, the factor $Z_\pm^{\alpha_r}$ of the new weight cancels, and every term of Eq.~(\ref{eq:GS1}) is multiplied by $Z^{-\alpha_r}|\det F|^{2r(1-\alpha_r)/d}$. Hence $S_{\alpha_r}(\Phi_F(P)|\mu_G)=S_{\alpha_r}(P|\mu_G)-r\ln Z+\frac{2r}{d}\ln|\det F|$.

\textit{Maximizer (Theorem~\ref{thm:max}).} By (i), $F=\rho^{-1/2}$ is a filter, with $\tr(F^\dagger F\rho)=d$ and $|\det F|^{2r/d}=(\det\rho)^{-r/d}$. Then $S_{\alpha_r}(P|\mu_G)=S_{\alpha_r}(\Phi_F(P)|\mu_G)+\frac rd\ln\det(d\rho)$, and $S_{\alpha_r}(\cdot|\mu_G)\le0$, with equality only at $\mu_G$, so the maximizer is $\Phi_{\sqrt\rho}(\mu_G)$.

\textit{Covariant rule (Observation~\ref{obs:rule}).} The Gaussian unitaries are filters that fix $I/d$, and they act transitively on each $\mathcal M_\pm$, so $\mathcal E(I/d)=q_+\mu_++q_-\mu_-$. By (iii), its average is $I/d$ only if $q_\pm=\frac12$. Every full-rank Gaussian $\rho$ is $I/d$ filtered by $\sqrt\rho$, which gives $\Phi_{\sqrt\rho}(\mu_G)$. Conversely, by (i), $F\sqrt\rho=\sqrt{Z\rho'}\,U$, with $U$ a Gaussian unitary, which leaves $\mu_G$ fixed.

\textit{Uniqueness of the order (Theorem~\ref{thm:order}).} As in Appendix~\ref{app:MSCC}, the proof of Theorem~\ref{thm:order} \smcite{sm:TI} carries over with three replacements. (i) In place of $|\langle0|\psi\rangle|^2$ we use $x=\langle\psi|\hat n_1|\psi\rangle$, the occupation of the first mode. Under $\mu_\pm$ its density is proportional to $x^{N-2}(1-x)^{N-2}$ \smcite{sm:GS}, continuous and positive on $(0,1)$. An ensemble on $\mathcal M_+$ whose density depends only on $x$ has its average fixed by $\mathbb E[x]$, since the average commutes with the Gaussian unitaries of modes $2,\dots,N$, which act irreducibly, and inequivalently, on the subspaces of $A_+$ with $\hat n_1=0$ and $\hat n_1=1$.
(ii) For $F=f^{1-\hat n_1}$, with $0<f<1$, $t=\Vert F\psi\Vert^2=x+f^2(1-x)$, and the filtered occupation, $x/t$, depends only on $x$. (iii) The filtered density is Eq.~(\ref{eq:12}), with $|\det F|^{2r/d}=f^{r}$. We compare ensembles with density $p(x)$ on $\mathcal M_+$ and $1$ on $\mathcal M_-$, whose $\mathcal M_-$ parts cancel. The argument in \smcite{sm:TI} then gives $\varphi'(c_F\,t^{r+1})\,t=\kappa_0+\kappa_1t$, with $c_F=1/(Z|\det F|^{2r/d})=2/[(1+f^2)f^{r}]$, whose solutions are those of Theorem~\ref{thm:order} with $d\to r$, on intervals of $s$ that contain $1$ and grow to $(0,\infty)$ as $f\to0$.

\clearpage
\onecolumngrid
\begin{center}
{\large\textbf{Supplemental Material for:\\[0.4ex]
Maximum Rényi entropy principle for the Scrooge ensemble from local filtering}}\\[2ex]
Agustín Agote$^{1,2}$ and Augusto J. Roncaglia$^{1,2}$\\[1ex]
{\small\textit{$^1$Universidad de Buenos Aires, Facultad de Ciencias Exactas y Naturales,
Departamento de Física, Ciudad Universitaria, 1428 Buenos Aires, Argentina}\\
\textit{$^2$CONICET - Universidad de Buenos Aires, Instituto de Física de Buenos Aires (IFIBA),
Ciudad Universitaria, 1428 Buenos Aires, Argentina}}
\end{center}
\vspace{1ex}

In this Supplemental Material, we provide proofs supporting the statements of the main text. It is organized as follows. In Sec.~\ref{sm:CC}, we show that projected ensembles satisfy the consistency condition exactly, because filtering $A$ commutes with measuring $B$. In Sec.~\ref{sm:TI}, we prove that the Rényi entropy of order $d/(d+1)$ is the only covariant separable entropy (Theorem~\ref{thm:order} of the main text). In Sec.~\ref{sm:GW}, we extend the covariance law to filters that weight the states by a general power of the norm, and show how the covariant order follows the Born exponent. In Sec.~\ref{sm:OI}, we show that the maximizers of the other Rényi orders, Shannon included, are not covariant. In Sec.~\ref{sm:NF}, we consider noninvertible filters. In Sec.~\ref{sm:GT}, we collect the Gaussian integrals used in Appendix~\ref{app:FFEB}. In Sec.~\ref{sm:CQ}, we treat conserved charges and derive the sector-wise Scrooge ensemble from covariance. Finally, in Sec.~\ref{sm:GS}, we complete the proofs for Gaussian states of free fermions (Theorem~\ref{thm:fermions}): the change of variables, Eq.~(\ref{eq:12}) of the main text, the distribution of a mode occupation used in Appendix~\ref{app:GS}, and the analogues of Secs.~\ref{sm:CC} and~\ref{sm:OI}.

\smsec{Filtering $A$ commutes with measuring $B$ (consistency condition)}{sm:CC}

This section shows that projected ensembles satisfy the consistency condition exactly, because a filter acts on $A$ and the measurement on $B$, so the two operations commute. To turn this into a statement about ensembles we need a distribution of global states with a given reduced state $\rho$ on $A$. We use uniformly random purifications of $\rho$, which are what Haar-random global states with reduced state $\rho$ give (shown below), and show two things: filtering $A$ takes a uniformly random purification of $\rho$ to a uniformly random purification of the filtered state $\rho'$, and it transforms the projected ensemble by $\Phi_F$.

Let $\dim B\ge d$ and $\{|i\rangle\}$ be a basis of $A$. Every purification of a full-rank $\rho$ is written $|\Psi_V\rangle = \sum_i\sqrt\rho\,|i\rangle\otimes V|i\rangle$, with $V:\mathbb C^d\to B$ an isometry. We call the measure $\Pi_\rho$ obtained with Haar-random $V$ (the first $d$ columns of a Haar unitary on $B$) \textit{uniform}. Let $F$ be invertible, $Z=\tr(F^\dagger F\rho)$, and $\rho'=F\rho F^\dagger/Z$. Then:

\begin{enumerate}
\item $\Vert(F\otimes I)\Psi_V\Vert^2=Z$ for every $V$; \label{it:CC1}
\item $\Psi\mapsto(F\otimes I)\Psi/\sqrt Z$ is a bijection from the purifications of $\rho$ onto those of $\rho'$ that takes $\Pi_\rho$ to $\Pi_{\rho'}$; \label{it:CC2}
\item for each $\Psi$ and each basis of $B$, the projected ensemble of the image is $\Phi_F$ of the projected ensemble of $\Psi$. \label{it:CC3}
\end{enumerate}

\textit{Proof.} \ref{it:CC1} The norm is $\tr[(F\otimes I)|\Psi\rangle\langle\Psi|(F^\dagger\otimes I)]=\tr(F^\dagger F\rho)$, because the partial trace of $|\Psi\rangle\langle\Psi|$ over $B$ is $\rho$. \ref{it:CC2} By polar decomposition, $F\sqrt\rho=\sqrt{Z\rho'}\,U$ with $U$ unitary on $A$, because $(F\sqrt\rho)(F\sqrt\rho)^\dagger = Z\rho'$. With the identity $\sum_iU|i\rangle\otimes V|i\rangle = \sum_i|i\rangle\otimes VU^{T}|i\rangle$,
\beq
\frac{(F\otimes I)\Psi_V}{\sqrt Z}=\sum_i\sqrt{\rho'}\,|i\rangle\otimes VU^{T}|i\rangle=\Psi'_{VU^T}.
\nonumber
\eeq
If $V$ is Haar, so is $VU^T$ (the Haar measure on isometries is invariant under unitaries on the right), so $\Pi_\rho$ goes to $\Pi_{\rho'}$. The inverse is the filter $F^{-1}$. \ref{it:CC3} The rows are $(I\otimes\langle b|)(F\otimes I)\Psi=F\tilde\psi_b$, which is the commutation of filtering $A$ and measuring $B$. $\square$

For a fixed basis of $B$, let $\mathcal E_B(\rho)$ be the projected ensemble of a uniform purification of $\rho$. It is a random ensemble, because it depends on the isometry $V$. Item~\ref{it:CC2} says that filtering a uniform purification of $\rho$ gives a uniform purification of $\rho'$, and item~\ref{it:CC3} says that its projected ensemble is filtered along with it. Together,
\beq
\mathcal E_B(\rho')=\Phi_F\big(\mathcal E_B(\rho)\big),
\nonumber
\eeq
where both sides are random ensembles and the equality is in distribution. This is the covariance, Eq.~(\ref{eq:7}). The projected ensemble therefore satisfies the consistency condition exactly, for every $\dim B\ge d$, without any limit of large dimension.

Uniform purifications are what Haar-random global states give. Let $|\Psi\rangle$ be Haar on $A\otimes B$, with $\dim B=n\ge d$, and write it as a $d\times n$ matrix of coefficients, $|\Psi\rangle=\sum_{i,b}M_{ib}\,|i\rangle\otimes|b\rangle$, so that $\rho_A=MM^\dagger$. Conditioned on $\rho_A=\rho$, $|\Psi\rangle$ is distributed according to $\Pi_\rho$, for almost every $\rho$, by symmetry. A unitary $W$ on $B$ changes $M$ into $MW^T$, leaves the Haar measure invariant and does not change $\rho_A$. Hence the distribution of $|\Psi\rangle$ conditioned on $\rho_A=\rho$ is also invariant under these unitaries. The states with $\rho_A=\rho$ are exactly the purifications $\Psi_V$, with $M=\sqrt\rho\,V^T$, and the unitaries on $B$ take any of them to any other, because $W$ takes the isometry $V$ to $WV$, and any isometry can be reached in this way. So all purifications of $\rho$ are equally likely, which is what the uniform measure $\Pi_\rho$ means. (In the same way, a uniform point on a sphere, conditioned on its height, is uniform on the circle at that height.)

This also connects the projected ensemble with Scrooge. Every full-rank $\rho$ is obtained from the maximally mixed state with the filter $F=\sqrt\rho$, since $\sqrt\rho\,(I/d)\sqrt\rho/\tr(\rho/d)=\rho$. Taking $I/d$ in place of $\rho$ and $F=\sqrt\rho$ in the equation above,
\beq
\mathcal E_B(\rho)=\Phi_{\sqrt\rho}\big(\mathcal E_B(I/d)\big).
\nonumber
\eeq
So it is enough to know the projected ensemble of the maximally mixed state, since that of every other $\rho$ follows by filtering. For large $B$, $\mathcal E_B(I/d)$ is close to Haar~\cite{Cotler23,Choi23,Ho22,Ippoliti23}. Therefore $\mathcal E_B(\rho)$ is close to $\Phi_{\sqrt\rho}(\Haar)=\Scr[\rho]$, and by the previous paragraph so is the typical projected ensemble of a Haar-random global state with reduced state $\rho$.

\smsec{Proof of Theorem~\ref{thm:order}: the only covariant separable entropy}{sm:TI}

Lemma~\ref{lem:cov} and its proof consider only Rényi entropies. Here we make no assumption on the form of the entropy. A \textit{separable entropy} is $\mathcal H_\varphi(P)=\int d\mu\,\varphi(p)$, with $\varphi$ continuously differentiable on $(0,\infty)$. We call it \textit{covariant} if every filter preserves the ranking it induces among ensembles with the same average: if $\rho_{P_1}=\rho_{P_2}$ and $\mathcal H_\varphi(P_1)>\mathcal H_\varphi(P_2)$, then $\mathcal H_\varphi(\Phi_F(P_1))>\mathcal H_\varphi(\Phi_F(P_2))$ for every filter $F$. It is the condition of the main text with $\mathcal C=\{P_1,P_2\}$.
The inequality in the conclusion is strict (the reason is explained in the proof below). It suffices to require it for ensembles whose density lies between two positive constants, a class that filters preserve.

\textbf{Theorem~\ref{thm:order}.} Let $d\ge2$ and $\varphi$ be continuously differentiable. $\mathcal H_\varphi$ is covariant if and only if $\varphi(s)=a_0+a_1\,s+a_2\,s^{d/(d+1)}$. Since $\int d\mu\,p=1$, the terms $a_0$ and $a_1$ are constants, so \textit{the Rényi entropy of order $d/(d+1)$ is the only separable entropy covariant under filters}, up to monotone transformations.

A filter \textit{multiplies} the density at each state by a factor that depends on the state, $\Vert F\psi\Vert^{2(d+1)}$, and the measure by another, $\Vert F\psi\Vert^{-2d}$, Eqs.~(\ref{eq:3}) and (\ref{eq:FFEB1}). For the two factors to compensate each other, $\varphi$ has to turn multiplications into multiplications, and the only such functions are the powers, which take a factor $\lambda$ in the density to $\lambda^\alpha$ in the entropy, where it can cancel the Jacobian. Among the powers, only the one with $(d+1)\alpha=d$ does so. Any other $\varphi$ fails. The logarithm (Shannon, Burg), for example, turns the factor into an additive term $\int d\mu \,p\ln\Vert F\psi\Vert$, which depends on the ensemble.

\textit{Proof.} Lemma~\ref{lem:cov} shows that the Rényi entropy of order $d/(d+1)$ satisfies the condition. It remains to show that no other separable entropy does. Since the condition holds for all ensembles and filters, a $\varphi$ that satisfies it also satisfies it in any particular case. We look at a well-chosen case and see that there only $\varphi=a_0+a_1\,s+a_2\,s^{d/(d+1)}$ survives.

The case is a pair of ensembles with the same average and the same entropy for every $\varphi$. We use the fact that the filter keeps equal entropies equal. The filtered ensembles $\Phi_F(P_1)$ and $\Phi_F(P_2)$ have the same average, Eq.~(\ref{eq:2}). If the filter made the entropy of one of them larger than that of the other, the inverse filter $F^{-1}$, which is also a filter ($\Phi_{F^{-1}}\circ \Phi_F={\rm id}$), would by covariance make the entropy of $P_1$ strictly larger than that of $P_2$, or the opposite, contradicting their equality. The definition therefore asks for a strict inequality, since with $\ge$ two equal entropies could become different without contradiction.

A small family of ensembles and filters suffices. Let $x=|\langle0|\psi\rangle|^2\in[0,1]$. We look at ensembles whose density relative to Haar depends only on $x$, $p(x)$. Under Haar, $x$ has density $(d-1)(1-x)^{d-2}$, and we write $d\mu(x)=(d-1)(1-x)^{d-2}dx$ for the Haar measure in the variable $x$. The mean value of $x$ in $P$ is $\mathbb E_P[x]=\int d\mu \,x\,p=\langle0|\rho|0\rangle$, because $x=\langle0|\psi\rangle\langle\psi|0\rangle$ and the average of $|\psi\rangle\langle\psi|$ in $P$ is $\rho$. The average state is $\rho={\rm diag}(a,b,\dots,b)$, since $p(x)$ does not change under the unitaries $U$ that fix $|0\rangle$ up to a phase (Haar and $x$ are both invariant under them), so $\rho=U\rho U^\dagger$ for every such $U$, which removes the off-diagonal elements and makes the remaining $d-1$ diagonal entries equal. As $\tr\rho=a+(d-1)b=1$, the average state is fixed by a single number, $a=\int d\mu\,x\,p$.

We use the filters $F={\rm diag}(1,f,\dots,f)$, with $0<f<1$, which map this family into itself. The success probability of the filter on a state $\psi$ is
\beq
t(x)=\Vert F\psi\Vert^2=x+f^2(1-x).
\nonumber
\eeq
It is linear in $x$. By Eqs.~(\ref{eq:3}) and (\ref{eq:FFEB1}),
\beq
\mathcal H_\varphi(\Phi_F(P))=|\det F|^2\int d\mu \varphi(c_F\,t^{d+1}p)\,t^{-d},
\nonumber
\eeq
with $c_F=1/(Z|\det F|^2)$ and $Z=\int d\mu\, t\,p=\tr(F^\dagger F\rho)$. Since $Z$ depends only on the average state, so does $c_F$.

We now build two ensembles, $P_+$ and $P_-$, with the same average, that no separable entropy can distinguish but the filter can. Let $\mathcal J_1,\mathcal J_2\subset(0,1)$ be two disjoint regions of $x$ with the same Haar measure, $\mu(\mathcal J_1)=\mu(\mathcal J_2)=m$ (with $\mu(\mathcal J_1)=\int_{\mathcal J_1} d\mu$), and the same mean of $x$; for example, $\mathcal J_1$ an interval and $\mathcal J_2$ two intervals, one on each side of $\mathcal J_1$, as below. The ensemble $P_+$ has density $1+\epsilon$ on $\mathcal J_1$, $1-\epsilon$ on $\mathcal J_2$, and $1$ elsewhere, with $0<\epsilon<1$. The ensemble $P_-$ is the opposite ($1-\epsilon$ on $\mathcal J_1$, $1+\epsilon$ on $\mathcal J_2$). They have two properties:

\begin{enumerate}
\item They have the same entropy for any $\varphi$. The two densities take the same values over regions of the same size, only in different places, so $\mathcal H_\varphi(P_\pm)=(1-2m)\varphi(1)+m\,\varphi(1+\epsilon)+m\,\varphi(1-\epsilon)$.
\item Both have average $I/d$: $\int d\mu \, p_\pm=1$ because $\mathcal J_1$ and $\mathcal J_2$ have the same measure, and $\int d\mu\,x\,p_\pm=\int d\mu\,x=1/d$ because $x$ has the same mean on $\mathcal J_1$ and on $\mathcal J_2$.
\end{enumerate}

Since the filter must keep these two entropies equal, $\mathcal H_\varphi(\Phi_F(P_+)) = \mathcal H_\varphi(\Phi_F(P_-))$. Both averages are $I/d$, so $c_F=df^{-2(d-1)}/(1+(d-1)f^2)$ is the same for both, and the difference comes only from $\mathcal J_1$ and $\mathcal J_2$:
\beq
0=\mathcal H_\varphi(\Phi_F(P_+))-\mathcal H_\varphi(\Phi_F(P_-))=|\det F|^2\,m\,\big(\overline\Omega_{\mathcal J_1}-\overline\Omega_{\mathcal J_2}\big),\qquad 
\Omega(x)=t^{-d}\Big[\varphi\big(c_F\,t^{d+1}(1+\epsilon)\big)-\varphi\big(c_F\,t^{d+1}(1-\epsilon)\big)\Big],
\nonumber
\eeq
with $\overline\Omega_{\mathcal J_1},\overline\Omega_{\mathcal J_2}$ the means of $\Omega$ on $\mathcal J_1$ and on $\mathcal J_2$. Thus, if two regions of the same size have the same mean of $x$, they have the same mean of $\Omega$.

This forces $\Omega$ to be affine in $x$, that is, a straight line. In general, if $\Omega$ is continuous and any two regions of equal measure $m$ and equal mean $x$ always have the same mean of $\Omega$, then $\Omega$ is affine in $x$. (Without the condition on the mean of $x$ only constants would satisfy the hypothesis, while an affine $\Omega$ satisfies it because its mean over a region is $l+r\bar x$.) Indeed, let $x_1<x_2$ and $0<\lambda<1$. Take $\mathcal J_2$ as two short intervals, one around $x_1$ with measure $\lambda m$ and another around $x_2$ with measure $(1-\lambda)m$, and $\mathcal J_1$ as an interval of measure $m$ with the same mean of $x$. As $m\to0$, $\mathcal J_1$ shrinks around $\lambda x_1+(1-\lambda)x_2$, and since $\Omega$ is continuous, $\Omega(\lambda x_1+(1-\lambda)x_2)=\lambda\Omega(x_1)+(1-\lambda)\Omega(x_2)$, which is what affine means. The equality obtained after the filter says that $\Omega$ satisfies this hypothesis, for each $\epsilon$. Dividing by $2\epsilon$ and letting $\epsilon\to0$, $\Omega/2\epsilon\to c_F\,\omega$, with $ \omega(x)= \varphi'(c_F\,t^{d+1})\,t$, and a limit of affine functions is affine. Since $t$ is linear in $x$, $\varphi'(c_F\,t^{d+1})\,t=\kappa_0+\kappa_1\,t$ for $t\in[f^2,1]$.

It remains to solve for $\varphi$, which is a change of variable. With $s=c_F\,t^{d+1}$, $t=(s/c_F)^{1/(d+1)}$, and the equation above divided by $t$ gives $\varphi'(s)=\kappa_1+\kappa_2\,s^{-1/(d+1)}$, a constant plus a power of $s$. Integrating, $\varphi=a_0+a_1\,s+a_2\,s^{d/(d+1)}$, with $a_1=\kappa_1$. This holds for $s\in \mathcal I_f=[\,c_Ff^{2(d+1)},\,c_F\,]=[\,df^4/(1+(d-1)f^2),\,c_F\,]$. All the $\mathcal I_f$ contain the value 1 (the filtered density averages to 1), and a function $\kappa_1+\kappa_2s^{-1/(d+1)}$ is determined by its values on an interval, so the constants do not depend on $f$. As $f\to0$, the union of the $\mathcal I_f$ is $(0,\infty)$. Each filter thus determines $\varphi$ on one interval of $s$, and these intervals overlap and grow to cover all $s>0$. The converse follows from Lemma~\ref{lem:cov}, since for that $\varphi$ the functional $\mathcal H_\varphi$ is a monotone function of $S_{\alpha_d}$, which at fixed average only shifts. $\square$

\smsec{General weight (Theorem~\ref{thm:order})}{sm:GW}

If the filter weights by $\Vert F\psi\Vert^{2\eta}$, Eq.~(\ref{eq:3}) changes the exponent $2(d+1)$ to $2(d+\eta)$ and $Z$ to $Z_\eta=\mathbb E_P[\Vert F\psi\Vert^{2\eta}]$. The computation of Lemma~\ref{lem:cov} gives
\beq
S_{d/(d+\eta)}(\Phi^{(\eta)}_F(P))=S_{d/(d+\eta)}(P)-\frac{d}{\eta}\ln Z_\eta+2\ln|\det F|,
\nonumber
\eeq
so the covariant order is $d/(d+\eta)$, the same as in Theorem~\ref{thm:SR} with Born exponent $\nu=2\eta$ (so $c=d/\eta$). But the filtered average is $\mathbb E_P[\Vert F\psi\Vert^{2\eta-2}F\psi\psi^\dagger F^\dagger]/Z_\eta$, which depends only on $\rho_P$ if and only if $\eta=1$. For $\eta\neq1$, two ensembles with the same average give, in general, different filtered averages.

\smsec{Maximizers of other orders are not covariant (Observation~\ref{obs:rule})}{sm:OI}

Let $\mathcal R_\alpha$ be the rule ``maximizer of $S_\alpha$ at fixed average'' ($\mathcal R_1$: Shannon). They all give Haar at $I/d$. Filtering Haar with $F$ gives, by Eq.~(\ref{eq:3}), the density $\propto\langle\psi|\Theta|\psi\rangle^{-(d+1)}$ with $\Theta=(FF^\dagger)^{-1}$. If $\mathcal R_\alpha$ were covariant, this would be its maximizer for the average $FF^\dagger/\tr(FF^\dagger)$. Maximizing $S_\alpha$ is maximizing $\int d\mu\,p^\alpha$ subject to $\int d\mu\,p=1$ and $\int d\mu\,p\,|\psi\rangle\langle\psi|=\rho$. With Lagrange multipliers $\lambda_0$ for the normalization and a Hermitian matrix $\Lambda$ for the average, the variation with respect to $p(\psi)$ at a maximizer with positive density gives $\alpha p^{\alpha-1}=\lambda_0+\langle\psi|\Lambda|\psi\rangle = \langle\psi|M|\psi\rangle$, with $M=\Lambda+\lambda_0 I$, that is, $p\propto \langle\psi|M|\psi\rangle^{-1/(1-\alpha)}$. For Shannon, $\ln p=-\langle\psi|M|\psi\rangle+{\rm const}$. One would then need $\langle\psi|M|\psi\rangle = a\,\langle\psi|\Theta|\psi\rangle^{s}$ with $s=(d+1)(1-\alpha)$, or, for Shannon, $\langle\psi|M|\psi\rangle=(d+1)\ln\langle\psi|\Theta|\psi\rangle+\text{const}$.

Take $\Theta$ with two distinct eigenvalues, with eigenvectors $e_1,e_2$, and the states parametrized as $\psi_u=\cos\frac u2e_1+\sin\frac u2e_2$, $0\le u<2\pi$. For them $\langle\psi_u|\Theta|\psi_u\rangle=b_1+b_2\cos u$ with $0<|b_2|<b_1$, and for any Hermitian $M$, $\langle\psi_u|M|\psi_u\rangle=m_1+m_2\cos u+m_3\sin u$. The condition would then require $a\,(b_1+b_2\cos u)^s=m_1+m_2\cos u+m_3\sin u$ for all $u$. The left side does not change under $u\to-u$, so $m_3=0$, and $(b_1+b_2x)^s$ would have to be a linear function of $x=\cos u$ on $[-1,1]$. For $s\neq0$ this holds only if $s=1$, that is, $\alpha=\alpha_d$. Shannon is excluded because $\ln(b_1+b_2x)$ is not linear in $x$ either. No assumption on the existence of maximizers is needed, since covariance would make $\Phi_F(\Haar)$ itself the maximizer, and its density is bounded above and below by positive constants, so the variation above applies to it.

\smsec{Noninvertible filters (Definition~\ref{def:filter}, Observation~\ref{obs:rule}, Lemma~\ref{lem:cov})}{sm:NF}

The main text uses only invertible filters. We now consider filters that are not invertible, as for a postselected projective measurement.

Scrooge is still covariant. If $F$ is not invertible, $\Phi_F(\Scr[\rho])=\Scr[F\rho F^\dagger/\tr(F^\dagger F\rho)]$, with Scrooge of a singular state defined on its support, as the Born-weighted direction of $\mathcal{CN}(0,F\rho F^\dagger)$. The proof is that of the main text without the change of variables. If $\tilde\psi\sim\mathcal{CN}(0,\rho)$, then $F\tilde\psi\sim\mathcal{CN}(0,F\rho F^\dagger)$ for any $F$, and the two Born weights combine (it is Theorem~2 of Tumulka~\cite{Tumulka26}, for arbitrary bounded operators). For instance, the qubit Kraus operator ${\rm diag}(0,\tfrac{\sqrt3}{2})$ has rank 1 and always leaves $|1\rangle$, which is $\Scr[|1\rangle\langle1|]$.

The rule is still unique. To fix it on every full-rank $\rho$ the invertible filters suffice (Observation~\ref{obs:rule}). The noninvertible ones only add conditions, which Scrooge satisfies, and extend the rule to singular $\rho$.

The entropy is taken on the image of the filter. By singular value decomposition, $F=U\Sigma V^\dagger$, and every filter is an invertible filter, followed by a projector onto a subspace $L$ of dimension $d_L={\rm rank}F$, followed by another invertible filter (within $L$). The filtered ensemble lives in $\mathbb{CP}(L)$ and is singular relative to Haar on $\mathbb C^d$, so the entropy to use is the one relative to Haar on $L$, and the covariant order becomes $d_L/(d_L+1)$, since Lemma~\ref{lem:cov} holds within each dimension. For the projective factor we have no covariance law of the entropy like Eq.~(\ref{eq:5}), only that Scrooge goes to Scrooge. The extension by continuity (a noninvertible filter as a limit of invertible ones) is heuristic.

\smsec{Gaussian tools (Appendix~\ref{app:FFEB})}{sm:GT}

For $\Sigma^{-1}>0$ Hermitian, $\int d^{2d}\tilde\psi\,e^{-\tilde\psi^\dagger \Sigma^{-1}\tilde\psi}=\pi^d \det \Sigma$. In polar coordinates $\tilde\psi=w\psi$, $d^{2d}\tilde\psi=\frac{\pi^d}{\Gamma(d)}\,2w^{2d-1}dw\,d\mu(\psi)$, and
\beq
\int_0^\infty dw\,2w^{2d-1}e^{-aw^2}=\frac{\Gamma(d)}{a^d}.
\nonumber
\eeq
Equating the two ways of computing $\int d^{2d}\tilde\psi\, e^{-\tilde\psi^\dagger \Sigma^{-1}\tilde\psi}$ gives
\beq
\int d\mu\,\langle\psi|\Sigma^{-1}|\psi\rangle^{-d}=\det \Sigma.\tag{S1}
\label{eq:S1}
\eeq
\textit{Direction of a Gaussian.} If $\tilde\psi\sim\mathcal{CN}(0,\Sigma)$, integrating its density $\det(\pi\Sigma)^{-1}e^{-w^2\langle\psi|\Sigma^{-1}|\psi\rangle}$ over $w$ with the radial formula, the direction has density $(\det\Sigma)^{-1}\langle\psi|\Sigma^{-1}|\psi\rangle^{-d}$ relative to Haar, normalized by Eq.~(\ref{eq:S1}). With $\Sigma=I$ it is Haar, and $w$ is independent of the direction. This is all that Appendix~\ref{app:FFEB} uses.

\smsec{Conserved charges (Proposition~\ref{prop:charges})}{sm:CQ}

Consider a conserved charge, with $A=\bigoplus_QA_Q$ and $\dim A_Q=d_Q$. A measurement that reveals the charge leaves $A$ in a single sector, and we say that the ensemble has a definite charge sector. The physical filters are those that respect the charge, $F=\bigoplus_QF_Q$. An example is $e^{-\beta H/2}$ with $H$ charge-conserving, which applied to a random state gives the canonical thermal pure quantum state~\cite{Sugiura13}. \medskip

\textbf{\Prop{prop:charges}.} Let a rule assign to every full-rank block-diagonal $\rho$ a definite-sector ensemble with average $\rho$. The only such rule that is covariant under charge-respecting filters is the sector-wise Scrooge ensemble,
\beq
\mathcal E(\rho)=\sum_Q\tr(\rho_Q)\,\Scr\big[\rho_Q/\tr\rho_Q\big],\qquad\rho=\bigoplus_Q\rho_Q .
\tag{S2}
\label{eq:CQ}
\eeq
This is the charge-sector form of the generalized Scrooge ensembles of Refs.~\cite{Mark24,Chang25}, which assign a Scrooge ensemble to each measurement outcome. It appears there and in Ref.~\cite{Igelspacher26} as a special case, whereas here it is singled out by covariance.
The symmetry enters only through which filters are physical, and the entropy inherits it. Since the problem decouples by sector (see below), the covariant order is $d_Q/(d_Q+1)$ in each sector. There is therefore no single Rényi entropy for the whole ensemble, and sectors of different dimension are maximized with different orders.

Equation~(\ref{eq:CQ}) contains known ensembles. Take a Haar state of fixed total charge and measure $B$ in the computational basis. The charge found in $B$ then reveals the sector of $A$, and the projected ensemble approaches a direct sum of Haar ensembles, one per sector~\cite{Chang25}. This is Eq.~(\ref{eq:CQ}) with blocks proportional to the identity. If the measurement does not reveal the sector of $A$, a single Scrooge ensemble is expected instead~\cite{Chang25,Igelspacher26}. A block-diagonal $\rho$ is compatible with both. Since $\rho$ cannot decide between them, the choice depends on what the measurement reveals, that is, which ensembles are admitted.

\textit{Proof of Proposition~\ref{prop:charges}.} The proof is that of Observation~\ref{obs:rule}, sector by sector. The block unitaries $\bigoplus_QU_Q$ are charge-respecting filters that leave $I/d$ fixed, so $\mathcal E(I/d)$ is invariant under them. In each sector it is therefore Haar, the only invariant probability on $\mathbb{CP}(A_Q)$. Every $\rho=\bigoplus_Q\rho_Q>0$ is $I/d$ filtered by $\sqrt\rho=\bigoplus_Q\sqrt{\rho_Q}$, which acts within each sector and takes Haar on $A_Q$ to $\Scr[\rho_Q/\tr\rho_Q]$ (Observation~\ref{obs:rule} in $A_Q$). The sector weights need not be computed, because the weight of sector $Q$ is the trace of block $Q$ of the average, and the average is $\rho$. This gives Eq.~(\ref{eq:CQ}), which is covariant by the block-by-block polar decomposition. $\square$

\textit{Entropy by sectors.} The state $\rho$ fixes the weight $\tr\rho_Q$ and the average $\rho_Q/\tr\rho_Q$ of each sector. Any combination $\sum_Q\theta_QS_{\alpha_{d_Q}}$ with $\theta_Q>0$ is therefore maximized sector by sector, and Theorem~\ref{thm:max} in each $A_Q$ gives Eq.~(\ref{eq:CQ}). All these functionals have the same maximizer, so the maximum does not single out one of them. The order $d_Q/(d_Q+1)$ in each sector comes from covariance.

\smsec{Gaussian states of free fermions (Theorem~\ref{thm:fermions})}{sm:GS}

We use the notation of the main text and of Appendix~\ref{app:GS}. The modes of $A$ have annihilation operators $a_j$, $j=1,\dots,N$, with $N\ge2$, occupations $\hat n_j=a_j^\dagger a_j$, and Majorana operators $\gamma_{2j-1}=a_j+a_j^\dagger$ and $\gamma_{2j}=i(a_j^\dagger-a_j)$, so that $\hat n_j=\frac12(1+i\gamma_{2j-1}\gamma_{2j})$. The vacuum is $|0\rangle$, and $\mathcal M_\pm$, $\mu_\pm$ and $\Pi_\pm$ are as in the main text and Appendix~\ref{app:GS}.

\textit{Change of variables.} Let $T(\psi)=F\psi/\Vert F\psi\Vert$, as in Appendix~\ref{app:FFEB}. By fact (i) of Appendix~\ref{app:GS}, $F=\zeta e^{\hat X}U$, with $\zeta>0$, $\hat X$ Hermitian and $U$ a Gaussian unitary. The scalar $\zeta$ moves no state and cancels in Eq.~(\ref{eq:12}). Because the Gaussian unitaries preserve $\mu_\pm$ and the change of variables of a composition is the product of the two, with $\Vert F\psi\Vert=\zeta\Vert e^{\hat X}U\psi\Vert$ and $|\det F|=\zeta^d\det e^{\hat X}$, it suffices to take $F=e^{\hat X}$. A single Majorana is a unitary that maps $\mathcal M_+$ onto $\mathcal M_-$, $\mu_+$ to $\mu_-$, and Gaussian filters to Gaussian filters, so it also suffices to take the even sector.

Allowing for a constant, which the scalar $\zeta$ otherwise absorbs, we write a Hermitian quadratic operator as
\beq
\hat X=h_0+\sum_{ij}h_{ij}\,a_i^\dagger a_j+\sum_{ij}\big(\Delta_{ij}\,a_i^\dagger a_j^\dagger+{\rm h.c.}\big),
\nonumber
\eeq
with $h_0$ real, $h$ Hermitian and $\Delta$ antisymmetric. $\hat X$ generates the flow $\psi_\tau=e^{\tau\hat X}\psi/\Vert e^{\tau\hat X}\psi\Vert$ on $\mathcal M_+$, and we write ${\rm div}_\psi(\hat X)$ for the divergence of its generator at $\psi$ with respect to $\mu_+$.
By Liouville's formula, $\partial_\tau\ln[d\mu_+(\psi_\tau)/d\mu_+(\psi)]={\rm div}_{\psi_\tau}(\hat X)$. In real local coordinates $y$, with $d\mu_+=\varrho(y)\,dy$ and $\mathcal V$ the vector field that generates the flow, ${\rm div}_{\psi}(\hat X)=\varrho^{-1}\nabla\cdot(\varrho\mathcal V)=\nabla\cdot\mathcal V+\mathcal V\cdot\nabla\ln\varrho$.

The divergence is fixed by symmetry up to two constants. It is linear in $\hat X$, and equivariant, ${\rm div}_{V\psi}(V\hat XV^\dagger)={\rm div}_\psi(\hat X)$, for Gaussian unitaries $V$, which preserve $\mu_+$. Since they act transitively on $\mathcal M_+$, it suffices to know it at $|0\rangle$, where it is invariant under the passive unitaries, which fix $|0\rangle$ up to a phase. A passive unitary acts as a Bogoliubov transformation $Va_i^\dagger V^\dagger=\sum_ku_{ki}a_k^\dagger$, with $u\in U(N)$~\cite{Bravyi05}. Substituting in $\hat X$, it maps $h\mapsto uhu^\dagger$ and $\Delta\mapsto u\Delta u^T$, and leaves $h_0$ fixed. The linear functionals of $h$ invariant under $h\mapsto uhu^\dagger$ are multiples of $\tr\, h$, by Schur's lemma, and those of $\Delta$ invariant under $\Delta\mapsto u\Delta u^T$ vanish, because $u=e^{i\theta}I$ gives $\Delta\mapsto e^{2i\theta}\Delta$. Two independent invariants are $\langle0|\hat X|0\rangle=h_0$ and $\tr(\Pi_+\hat X)=\frac d2(h_0+\frac12\tr \, h)$, since each mode is occupied in half of the even states. Hence
\beq
{\rm div}_{|0\rangle}(\hat X)=\xi_1\,\tr(\Pi_+\hat X)+\xi_2\,\langle0|\hat X|0\rangle .
\nonumber
\eeq
Two flows fix the constants. The identity generates no flow, so $\frac d2\xi_1+\xi_2=0$. 
The total number operator, $\hat N=\sum_j\hat n_j$, acts on the Thouless coordinates~\cite{Thouless60,RingSchuck80}, $e^{\frac12\sum_{ij}z_{ij}a_i^\dagger a_j^\dagger}|0\rangle$ with $z$ complex antisymmetric, as $z\mapsto e^{2\tau}z$, because $e^{\tau\hat N}a_i^\dagger a_j^\dagger e^{-\tau\hat N}=e^{2\tau}a_i^\dagger a_j^\dagger$ and $e^{\tau\hat N}|0\rangle=|0\rangle$. This flow is generated by the vector field $\mathcal V(z)=2z$. Since $\mathcal V$ vanishes at $z=0$, the term $\mathcal V\cdot\nabla\ln\varrho$ vanishes there, leaving the ordinary divergence of $\mathcal V$ on $\mathbb C^{N(N-1)/2}\cong\mathbb R^{N(N-1)}$, namely $2N(N-1)$. With $\langle0|\hat N|0\rangle=0$ and $\tr(\Pi_+\hat N)=dN/4$, this gives $\xi_1=4r/d$ and $\xi_2=-2r$. Integrating along the flow from $\tau=0$ to $1$, with $\partial_\tau\ln\Vert e^{\tau\hat X}\psi\Vert^2=2\langle\psi_\tau|\hat X|\psi_\tau\rangle$,
\beq
\ln\frac{d\mu_+(T(\psi))}{d\mu_+(\psi)}=\frac{4r}{d}\tr(\Pi_+\hat X)-r\ln\Vert e^{\hat X}\psi\Vert^2 .
\nonumber
\eeq
By fact (ii) of Appendix~\ref{app:GS}, $\det e^{\hat X}=e^{2\tr(\Pi_+\hat X)}$, so $d\mu_\pm(\psi')=|\det F|^{2r/d}\,\Vert F\psi\Vert^{-2r}\,d\mu_\pm(\psi)$. Together with the Born weight $\Vert F\psi\Vert^2/Z$, this gives Eq.~(\ref{eq:12}).

\textit{Distribution of a mode occupation.} Write $x=\langle\psi|\hat n_1|\psi\rangle=\frac12(1+v_2)$, with $v\in\mathbb R^{2N}$ given by $v_a=i\langle\psi|\gamma_1\gamma_a|\psi\rangle$ for $a\ne1$ and $v_1=0$. For a pure Gaussian state, $v$ is a row of its correlation matrix, and hence a unit vector~\cite{Bravyi05,SuraceTagliacozzo22}, orthogonal to $e_1$. A Gaussian unitary $V$ rotates the Majoranas, $V^\dagger\gamma_aV=\sum_bO_{ab}\gamma_b$ with $O\in SO(2N)$~\cite{Bravyi05}, and if $O$ fixes $e_1$, the state $V\psi$ has vector $Ov$. These unitaries preserve $\mu_\pm$, and their rotations act transitively on the unit sphere of the subspace orthogonal to $e_1$, $S^{2N-2}$. Under $\mu_\pm$, $v$ is therefore uniform on $S^{2N-2}$, and its component $v_2$ has density proportional to $(1-v_2^2)^{N-2}$. This is the density of $x$ used in Appendix~\ref{app:GS}.

\textit{Other orders.} As in Sec.~\ref{sm:OI}, the maximizers of the other orders are not covariant. On $\mathcal M_\pm$, stationarity gives $\alpha p^{\alpha-1}=\langle\psi|M|\psi\rangle$, and the filtered $\mu_G$ has density proportional to $\langle\psi|\Theta|\psi\rangle^{-(r+1)}$, with $\Theta=(FF^\dagger)^{-1}$, so that $s=(r+1)(1-\alpha)$. The one-parameter family of Sec.~\ref{sm:OI} must now consist of Gaussian states. Take $F=e^{\epsilon(\hat n_1+\hat n_2)}$, with $\epsilon\ne0$. The eigenvectors $|0\rangle$ and $a_1^\dagger a_2^\dagger|0\rangle$ of $\Theta$ have distinct eigenvalues, and $\psi_u=\cos\frac u2|0\rangle+\sin\frac u2\,a_1^\dagger a_2^\dagger|0\rangle$ is proportional to $e^{\tan(u/2)a_1^\dagger a_2^\dagger}|0\rangle$, which is Gaussian, for every $u$. The rest of Sec.~\ref{sm:OI} is unchanged and gives $s=1$, that is, $\alpha=\alpha_r$.

\textit{Projected ensembles.} Let $B$ have $N_B\ge N$ modes, with Majoranas $\gamma^B_a$, measured in a Gaussian basis such as mode occupations, and fix the global parity. Write $\gamma^A_a$ for the Majoranas of $A$, and let $\Psi_0$ be the state with $i\gamma^A_a\gamma^B_a\Psi_0=\Psi_0$ for $a=1,\dots,2N$ and the remaining modes of $B$ empty. This is the Fock state in which the $2N$ modes $c_a = \frac12(\gamma^A_a+i\gamma^B_a)$ are occupied and the remaining modes of $B$ are empty, so it exists, is unique up to a phase, and is Gaussian. It is the fermionic analogue of the maximally entangled state of Sec.~\ref{sm:CC}, and its reduced state on $A$ is $I/d$. From $\gamma^A_a\Psi_0=i\gamma^B_a\Psi_0$ it follows that $\gamma^A_a\gamma^A_b\Psi_0=\gamma^B_b\gamma^B_a\Psi_0$, and, since bilinears on $A$ commute with bilinears on $B$,
\beq
e^{\frac12\sum_{ab}X_{ab}\gamma^A_a\gamma^A_b}\,\Psi_0=e^{-\frac12\sum_{ab}X_{ab}\gamma^B_a\gamma^B_b}\,\Psi_0
\nonumber
\eeq
for every complex antisymmetric $X$. A Gaussian operator on $A$ can thus be moved to $B$. This replaces the identity $\sum_iU|i\rangle\otimes V|i\rangle=\sum_i|i\rangle\otimes VU^T|i\rangle$ of Sec.~\ref{sm:CC}. 

The Gaussian purifications of a full-rank Gaussian $\rho$, with the parity of $\Psi_0$, are $(\sqrt{d\rho}\otimes U_B)\Psi_0$, with $U_B$ a Gaussian unitary on $B$, a special case of the modewise decomposition of pure Gaussian states~\cite{BoteroReznik04}. Indeed, if $\Psi$ is such a purification, $\Psi'=((d\rho)^{-1/2}\otimes I)\Psi$ is normalized, Gaussian by fact (i) of Appendix~\ref{app:GS}, and has reduced state $I/d$ on $A$, so its correlation matrix $\Gamma_{ab}=i\langle\gamma_a\gamma_b\rangle$ ($a\ne b$) vanishes within $A$, $\Gamma=\left(\begin{smallmatrix}0 & C \\ -C^T & \mathcal B\end{smallmatrix}\right)$, with $C$ the correlations between $A$ and $B$ and $\mathcal B$ those within $B$. Purity, $\Gamma^2=-I$~\cite{Bravyi05}, reads, block by block,
\beq
CC^T=I,\qquad C\mathcal B=0,\qquad \mathcal B^2=-(I-C^TC).
\nonumber
\eeq
By the first relation, the rows of $C$ are orthonormal, so $\tilde\gamma^B_a=\sum_bC_{ab}\gamma^B_b$, $a=1,\dots,2N$, are Majoranas of $B$, $\{\tilde\gamma^B_a,\tilde\gamma^B_b\}=2\delta_{ab}$. They satisfy $i\gamma^A_a\tilde\gamma^B_a\Psi'=\Psi'$, since this operator has eigenvalues $\pm1$ and expectation $\sum_bC_{ab}\,i\langle\gamma^A_a\gamma^B_b\rangle=\sum_bC_{ab}^2=1$. The remaining Majoranas of $B$, orthogonal to the $\tilde\gamma^B_a$, span the range of the projector $I-C^TC$. By the second relation and $\mathcal B^T=-\mathcal B$, $\mathcal B=(I-C^TC)\,\mathcal B\,(I-C^TC)$, so the correlations within $B$ involve only the remaining Majoranas, which are uncorrelated with the $\tilde\gamma^B_a$, and, since $C(I-C^TC)=0$, also with $A$. By the third relation, their correlation matrix squares to $-I$, so they are in a pure Gaussian state. A unitary $U_B$ with $U_B\gamma^B_aU_B^\dagger=\tilde\gamma^B_a$ that maps the vacuum of the remaining modes of $B$ to that state takes $\Psi_0$ to $\Psi'$, up to a phase. It implements an orthogonal map of the Majoranas of $B$, and preserves parity because $\Psi_0$ and $\Psi'$ have the same parity, so it is a Gaussian unitary.

We call the measure with $U_B$ Haar-distributed on the Gaussian unitaries of $B$ uniform. Items~\ref{it:CC1}--\ref{it:CC3} of Sec.~\ref{sm:CC} then hold with the polar decomposition of Appendix~\ref{app:GS}, fact (i), in place of that of $F\sqrt\rho$, and, by the same symmetry argument, Haar-random Gaussian global states conditioned on $\rho_A=\rho$ are uniform Gaussian purifications. Hence $\mathcal E_B(\rho')=\Phi_F(\mathcal E_B(\rho))$ for every Gaussian filter, exactly. For large $N_B$, the projected ensemble of a typical random Gaussian state with $\rho_A=I/d$ approaches $\mu_G$~\cite{Bejan25}, so, as in Sec.~\ref{sm:CC}, the typical projected ensemble at $\rho$ approaches $\Phi_{\sqrt\rho}(\mu_G)$, the fermionic Gaussian Scrooge ensemble.

\end{document}